\documentclass{elsart4-1}
\usepackage[applemac]{inputenc}
\usepackage{natbib}

\usepackage{amssymb}

\usepackage[english,francais]{babel}

\newtheorem{e-proposition}[theorem]{Proposition}

\newtheorem{e-definition}[theorem]{Definition\rm}

\renewcommand{\theequation}{\arabic{equation}}
\def\og{\leavevmode\raise.3ex\hbox{$\scriptscriptstyle\langle\!\langle$~}}
\def\fg{\leavevmode\raise.3ex\hbox{~$\!\scriptscriptstyle\,\rangle\!\rangle$}}

\def\cras{CRAS}
\def\lev{Le~Verrier }
\def\bq{\medskip\hskip 1cm\begin{minipage}{12cm} \normalsize} 
\def\eq{
\end{minipage}\\ \medskip\noindent} 
\def\bqs{\medskip\hskip 1cm\begin{minipage}{12cm} \normalsize\it} 

\begin{document}

\centerline{Physics or Astrophysics/Header}
\begin{frontmatter}


\selectlanguage{english}
\title{Des premiers travaux de Le Verrier à la découverte de Neptune.}


\selectlanguage{english}
\author[]{Jacques Laskar},
\ead{laskar@imcce.fr}

\address[]{ASD/IMCCE, CNRS-UMR8028, Observatoire de Paris,  PSL Research University, UPMC, 77 Avenue Denfert-Rochereau, 75014 Paris, France}


\medskip
\begin{center}
{\small Received *****; accepted after revision +++++}
\end{center}

\begin{abstract}
{\bf From Le Verrier's first works to the discovery of Neptune.}
Urbain-Jean-Joseph Le Verrier was born in Saint-L\^o on March 11, 1811. 
He entered the Ecole Polytechnique in 1831, from which he was to emerge 8th two years later.
After first devoting himself to chemistry, in 1836 he obtained a position as an astronomy assistant at the Ecole Polytechnique.
This choice will decide his future career which culminates with the discovery of Neptune in 1846.
Le Verrier wrote more than 200 contributions in the \cras. 
These contributions are very varied: some original articles but also reports on publications
 published elsewhere, sometimes even simple notes of a single page. The whole set gives a very vivid vision
of the development of the science of the XIXth century. At that time, the Comptes Rendus are really a 
reflection of the debates of the sessions of the Academy.
They are published very quickly, and leave a large freedom of speech to the authors. They are therefore
 a snapshot of  the sometimes lively polemics which animated the sessions of the Academy of Sciences. In this limited essay,
we will mainly look to 
the first years of the career of Le Verrier until the discovery of Neptune.

{\it To cite this article: J. Laskar C. R.
Physique 6 (2017).}

\vskip 0.5\baselineskip

\selectlanguage{francais}
\noindent{\bf R\'esum\'e}
\vskip 0.5\baselineskip
\noindent
Urbain-Jean-Joseph Le Verrier est né à Saint-L\^o le 11 mars 1811. 
Il entre à l'Ecole Polytechnique en 1831 d'où il sortira 8ème deux ans après. 
Après s'être d'abord consacré à la chimie, il obtient en 1836 un poste de répétiteur en astronomie à l'Ecole Polytechnique.
Ce choix décidera de sa carrière future qui culminera avec la découverte de Neptune en 1846.
Le Verrier a écrit plus de 200 contributions dans les \cras. 
Ces contributions sont très variées : il y a certes de véritables articles originaux mais aussi des rapports sur des publications 
publiées  ailleurs, parfois de simples notes d'une seule page. L'ensemble donne une vision très vivante
du développement de la science du XIXème siècle. En ce temps-là, les Comptes Rendus sont  vraiment le reflet des débats des séances
de l'Académie.
Ils sont publiés très rapidement, et laissent une très grande liberté de parole aux auteurs. On y retrouve  
de manière instantanée les polémiques parfois vives qui ont animé les séances de l'Académie des sciences. Dans  cet essai limité
nous survolerons principalement
les premières années de la carrière de Le Verrier jusqu'à la découverte de Neptune.

{\it Pour citer cet article~: J. Laskar, C. R.
Physique 6 (2017).}

\keyword{Keyword1; Keyword2; Keyword3 } \vskip 0.5\baselineskip
\noindent{\small{\it Mots-cl\'es~:} Mot-cl\'e1~; Mot-cl\'e2~;
Mot-cl\'e3}}
\end{abstract}
\end{frontmatter}

\selectlanguage{francais}

\selectlanguage{english}
\section{Introduction}
Urbain-Jean-Joseph Le Verrier est né à Saint-L\^o le 11 mars 1811. D'abord élève au collège communal de Saint-L\^o, puis 
au collège royal de Caen, il entre à l'Ecole Polytechnique en 1831 d'où il en sortira 8ème deux ans après. 
Après s'être d'abord consacré à la chimie, il obtient en 1836 un poste de répétiteur en astronomie à l'Ecole Polytechnique.
Ce choix décidera de sa carrière future qui culminera avec la découverte de Neptune en 1846 \citep[e.g.][]{Lequ2009a}.
Le Verrier a écrit plus de 200 contributions dans les \cras. On en trouvera une liste extensive dans le livre 
édité par l'Institut de France à l'occasion du centenaire de la naissance de Le Verrier \citep{Inst1911a}.
Ces contributions sont très variées : il y a certes de véritables articles originaux mais aussi des rapports sur des publications 
plus importantes publiées  ailleurs, des présentations d'articles pour lesquels l'aval de l'Académie des sciences est demandé
afin de  pouvoir les publier. Ce sera le cas de plusieurs  travaux de Le Verrier qui paraîtront dans le Journal de Liouville.
Ces contributions sont aussi parfois de simples notes d'une seule page. L'ensemble donne une vision très vivante
du développement de la science de l'époque. En ce temps-là, les Comptes Rendus sont  vraiment le reflet des débats des séances
de l'Académie des sciences.
Ils sont publiés très rapidement, et laissent une très grande liberté de parole aux auteurs. On y retrouve donc
de manière instantanée les polémiques parfois vives qui ont animé les séances de l'Académie des sciences. Dans  cet essai limité
nous survolerons  
les premières années de la carrière de Le Verrier jusqu'à la découverte de Neptune.
\label{}

\section{Premiers travaux, premières polémiques}
En 1839, \lev  a 28 ans. Pour son premier travail en astronomie, il s'attaque à un problème fondamental,
 lié à la question de la stabilité du système solaire. À la fin du XVIIIème Laplace et Lagrange 
avaient résolu ce problème, au moins dans  une approximation linéaire. Laplace, montre que les demi-grands
axes des planètes sont invariants dans un système moyen du premier ordre \citep{Lapl1776a}. Par ailleurs, Lagrange 
développe la méthode qui permet de calculer les variations à long terme des orbites des planètes. Il l'applique en un premier temps 
aux seules inclinaisons \citep{Lagr1778a}, en soumettant son manuscrit à l'Académie des Sciences en 1774
où il tombe entre les mains de Laplace qui s'empresse 
d'appliquer la même méthode aux mouvements des périhélies et  des excentricités \citep{Lapl1775a} 
(pour une description plus détaillée de ces épisodes, voir  \citep{Lask1992a,Lask2013a}). Quelques années plus tard, Lagrange 
construit la première solution complète  de l'évolution à long terme des orbites des planètes du Système solaire
\citep{Lagr1781a,Lagr1782a}. La solution de Lagrange ne comprenait que les planètes visibles à l'\oe il nu
(Mercure, Vénus, la Terre, Mars, Jupiter et Saturne). 
Le Verrier se propose  alors  d'étendre  cette solution en y incluant Uranus, découverte par Herschel en 1781. 
Des extraits de ses travaux seront présentés par François Arago, Félix Savary et Joseph Liouville dans 
les séances de l'Académie du 16 septembre et 14 octobre 1839.
Le Verrier tient  à montrer l'originalité de son travail et à affirmer son importance, en relativisant 
les résultats de Lagrange qui, selon lui, ne disposait pas de valeurs de masses des planètes assez précises. 

\bqs
Cette détermination a été effectuée par Lagrange,
dans les Mémoires de l'Académie de Berlin, pour 1782. Mais les formules
qu'il donne pour les quatre planètes dont nous parlons sont complètement
inexactes, et leur emploi doit être rejeté \citep{Le-V1839a}.
\eq

Il n'était par ailleurs pas le premier à entreprendre ce travail. Gustave de Pontécoulant, dans sa {\it Théorie analytique du système du monde} 
s'y était déjà employé, \citep{de-P1856c}, mais avait commis plusieurs erreurs que Le Verrier s'empresse de signaler lors de la présentation de 
son travail à l'Académie des Sciences 

\bqs
 M. de Pontécoulant a repris
ce travail dans le troisième volume du Système analytique du monde;
mais les nombres qu'il y a consignés sont tous affectés des erreurs les plus
graves et qui dépassent, quelquefois, jusqu'à 20000 et même 40000 fois les
valeurs absolues de ces nombres ; aussi en supposant le temps nul dans ses
formules, trouve-t-on des valeurs des excentricités et des positions de périhélies
qui n'ont pas le plus léger rapport avec celles que l'observation détermine \citep{Le-V1839a}.
\eq

De Pontécoulant rétorque par une lettre à l'Académie qui ne dément pas les attaques de Le Verrier. 

\bqs
Quant au travail que j'ai exécuté, j'espère que l'Académie reconnaîtra le zèle avec lequel je m'y suis livré, 
et qu'elle jugera qu'à défaut d'autre mérite, il aurait encore celui d'avoir mis en évidence les véritables difficultés 
de la question, et d'avoir fourni aux géomètres l'occasion de chercher à les surmonter 
ou d'imaginer des méthodes qui en soient exemptes pour arriver au but que je m'étais proposé \citep{De-P1839a}.
\eq

Pour enfoncer le clou, Le Verrier envoie une nouvelle note qui sera publiée dans les comptes rendus de la même  séance
du 28 octobre dans laquelle il démontre, chiffres à l'appui, les erreurs de G. de Pontécoulant \citep{Le-V1839c}. Cela n'arrange surement pas les affaires de ce dernier 
qui venait d'échouer le 3 juin 1839  à une élection à l'Académie, au profit de Liouville. Déjà membre de l'Académie de Berlin, et de la Royal Society, il ne sera jamais élu à l'Académie des Sciences. Les attaques de Le Verrier trouveront un écho auprès de Liouville 
\citep{Liou1840b} et même d'Arago \citep{Arag1840a,de-P1840a}.

Le rapport sur le travail de Le Verrier sera présenté par Liouville dans la séance du 30 mars 1839. Il fait état de la difficulté de 
considérer un problème à 7 planètes, car la solution exige le calcul des valeurs propres d'une matrice $7\times 7$ 
et donc de résoudre une équation du 7ème degré. Lagrange avait contourné le problème en séparant le système de Jupiter et Saturne 
du reste, ce qui n'est plus possible lorsque Uranus est pris en compte. 

\bqs
question très compliquée lorsque l'on considère
à la fois les sept planètes principales: le calcul pénible qu'elle exige et que
personne jusqu'ici n'avait effectué d'une manière exacte, M. Le Verrier l'a
entrepris avec succès dans le Mémoire dont nous rendons compte aujourd'hui \citep{Liou1840a}.
\eq 

Les  résultats de Le Verrier seront publiés dans le Journal de Liouville \citep{Le-V1840e,Le-V1840f} et une version plus longue 
incluant les tables dans les {\it Additions à  la Connaissance des Temps} \citep{Le-V1840g}.
Le Verrier ne fait pas qu'étendre la solution séculaire à un système à 7 planètes comprenant Uranus. Il tient aussi 
à ce que son travail reste valide  en dépit des incertitudes sur les valeurs des masses planétaires. Il va donc 
non seulement donner la solution nominale, mais aussi ses dérivées partielles par rapport aux variations des masses planétaires. 

Dans la même année, il recherche des méthodes pour calculer de manière effective les développements de la fonction perturbatrice entre deux planètes, non pas en séries entières des petits paramètres que sont les excentricités et les inclinaisons, mais en séries de Fourier 
à deux arguments (les longitudes moyennes des deux planètes) \citep{Le-V1840b,Le-V1840c,Le-V1841c}. Il utilisera ces travaux pour arriver 
à modéliser correctement le mouvement de la petite planète Pallas dont l'excentricité est importante (0.23) en tenant compte d'un terme 
quasi résonant d'ordre élevé ( $18\lambda_5 - 7 \lambda$) entre les longitudes moyennes de Jupiter ($\lambda_5$) et Pallas ($\lambda$)
\citep{Le-V1841b,Le-V1843c}. Plus important encore, Le Verrier met au point, à travers ces calculs, les méthodes qu'il utilisera plus tard pour l'étude des perturbations d'Uranus.  

Il veut ensuite étendre le calcul du système séculaire des planètes aux ordres supérieurs. Lagrange et Laplace n'avaient 
considéré que l'approximation linéaire, tout comme Le Verrier dans son premier travail \citep{Le-V1840e,Le-V1840f,Le-V1840g}.
Comment savoir si en développant à des  degrés plus élevés par rapport aux petits paramètres (les excentricités et les inclinaisons), 
 on ne va pas modifier de manière substantielle les résultats obtenus dans le cadre linéaire, à un point tel que la stabilité du système 
 solaire sera remise en cause ? 
 
 \bqs
 Je me suis proposé de reconnaître si, par la méthode des approximations successives, les intégrales se développent effectivement en séries assez convergentes pour qu'on puisse répondre de la stabilité du système planétaire \citep{Le-V1840d,Le-V1841d}.
 \eq
 
 Après un calcul difficile des termes du troisième degré  dans le système des  planètes géantes  (Jupiter, Saturne, Uranus), 
 Le Verrier constate que leur contribution principale est un léger changement des fréquences du système (les fréquences séculaires).
 Il en concluera que les termes de degrés supérieurs ne donneront que des variations encore plus faibles, 
et que la seule amélioration possible consisterait dans une meilleure connaissance des masses planétaires. 
 
 \bqs
 La considération des termes du troisième ordre était donc nécessaire
pour donner aux  formules des inégalités séculaires de Jupiter, Saturne et
Uranus, toute l'exactitude dont elles sont susceptibles. Cette exactitude,
sous le point de vue astronomique, ne pourra plus être dépassée que lorsque,
par leurs développements observés avec soin pendant une longue
suite d'années, les inégalités de ces planètes auront donné le moyen d'estimer
leurs masses d'une manière plus rigoureuse qu'on ne peut le faire actuellement \citep{Le-V1840d,Le-V1841d}.
 \eq   
 
Il ne se doute pas que G. Hill montrera plus tard que les termes du second ordre des masses qu'il a  négligé donnent, 
à cause de la quasi-résonance entre Jupiter et Saturne, une contribution beaucoup plus importante 
aux variations de fréquences que la prise en considération  des termes de degré 3. En effet, la fréquence principale 
$g_5$ du périhélie de Jupiter est de $22.4273$"/an dans \citep{Le-V1840g}. La correction de degré 3 
de  \citep{Le-V1840d,Le-V1841d} est de $0.3231$"/an alors qu'en tenant compte de  
la contribution au second ordre des masses, Hill trouve $g_5 =27.7301$"/an \citep{Hill1897a}, beaucoup plus  proche de
la valeur moderne de $28.2450$"/an  \citep{LaskRobu2004a}.

Pour ce qui concerne les planètes terrestres (Mercure, Vénus, la Terre et Mars), la tâche est plus difficile, et Le Verrier 
voit apparaître le problème des petits diviseurs qui compromettent la stabilité des résultats des approximations successives.
Il se rend compte aussi qu'une très faible variation des masses des planètes, encore incertaines à l'époque pour celles qui comme 
Mercure ou Vénus n'ont pas de satellites, peut donner une valeur nulle à un diviseur, ou le faire changer de signe. Il en conclut alors 

\bqs
Il paraît donc impossible, par la méthode des approximations successives,
de prononcer si, en vertu des termes de la seconde approximation,
le système composé de Mercure, Vénus, la Terre et Mars jouira d'une
stabilité indéfinie; et l'on doit désirer que les géomètres, par l'intégration rigoureuse
des équations différentielles, donnent les moyens de lever cette
difficulté, qui peut très bien ne tenir qu'à la forme \citep{Le-V1840d,Le-V1841d}.
\eq

Cet appel aux {\it géomètres} sera entendu par  Henri Poincaré qui montrera une cinquantaine  d'années plus tard, que les difficultés 
évoquées par Le Verrier ne sont pas dues à la forme, mais sont intrinsèques aux problèmes planétaires, 
qui, en général, ne sont  pas intégrables. Pour plus de détails, on pourra consulter   \citep{Lask2013a}  et les références citées.

\section{La théorie de Mercure} 
Après les mouvements séculaires des planètes, Le Verrier s'attaque à un autre problème ardu : le mouvement de Mercure. 
Par des méthodes de perturbations, il raffine les solutions mais ne s'arrête pas là, car il procède aussi à l'ajustement 
des paramètres de la solution par rapport aux observations disponibles, dont celles, de bonne qualité, 
obtenues sur les lunettes méridiennes  de l'Observatoire de Paris.

\bqs
Je dois à la libéralité scientifique de l'illustre directeur de notre Observatoire,
M. Arago, d'avoir pu puiser sans réserve dans ces précieux recueils,
encore inédits. J'ai fait tous mes efforts pour que l'exactitude de la théorie ne
reste pas au-dessous de la précision des observations qui m'étaient confiées \citep{Le-V1843a}.
\eq

Il utilisera aussi pour cette première solution de l'orbite de Mercure, les passages de Mercure devant le soleil 
des années 1697, 1723, 1736, 1743, 1753, 1769, 1782, 1786, 1789, 1799, 1802 et 1832. Le mémoire est analysé 
par une commission composée de Arago, Mathieu, Damoiseau, et Liouville, et présenté avec des vérifications 
effectuées par Laugier à la lunette méridienne de l'Observatoire de Paris \citep{Le-V1843b}. Il sera publié dans le Journal de Liouville 
\citep{Le-V1843g}. Ces Tables de Mercure sont aussi l'occasion pour Le Verrier de perfectionner ses techniques 
de perturbations planétaires. Grâce à ses perturbations sur l'orbite de Mercure, il détermine la masse de Vénus 
qu'il trouve égale à (1/390 000) masse solaire, peu différente de la valeur obtenue par Burckhardt 
en considérant les variations séculaires de l'obliquité (1/401 847), et en accord avec la valeur actuelle (1/408 524 masse solaire).
Il continuera à s'intéresser au mouvement de Mercure et à ses observations \citep{Le-V1843d},  
et cette théorie de Mercure lui servira d'étalon pour ses travaux futurs \citep{Le-V1843h}. 
Le rapport sur ces travaux est établi par Laugier, au sein d'une commission comprenant également  Damoiseau et Liouville. 
Il est présenté de manière très positive  dans la séance du 8 août 1845 \citep{Laug1845a}, en préconisant de publier 
ce travail dans le {\it Recueil des Savants étrangers}, c'est-à-dire des savants non-académiciens. Ces commissions 
effectuaient donc de manière extrêmement sérieuse, ce qui est demandé de nos jours aux rapporteurs des revues scientifiques. 
L'avantage de cette formule pour l'histoire des sciences étant que l'intégralité du rapport était publiée dans les CRAS.
En réalité, Le Verrier préfèrera publier son ouvrage dans la {\it Connaissance des Temps} qui lui permettait sans doute 
des délais de publication beaucoup plus réduits \citep{Le-V1845g}.

Deux ans  après que Le Verrier eut présenté à l'Académie sa "Théorie du mouvement de Mercure", un passage 
de Mercure devant le soleil doit avoir lieu, le 8 mai 1845. Le Verrier s'empresse de faire 
ses propres prédictions, en utilisant sa nouvelle solution, qui seront présentées lors de la séance du 3 mars 1845
\citep{Le-V1845a}. Il est impressionnant de voir que le compte rendu des premières observations du passage de Mercure 
devant le soleil est présenté en séance dès le 26 mai 1845 \citep{Le-V1845d}, Le Verrier y vérifie l'excellence de ses prédictions, 
avec seulement 18 secondes d'écart avec l'observation du premier contact interne, contre 77 secondes pour les éphémérides de Berlin. 
Une deuxième série d'observations effectuées à Cincinnati est  présentée  le 29 septembre 1845. En raison de l'incertitude de 
la longitude de l'Observatoire de Cincinnati, Le Verrier préfère cette fois-ci comparer les durées totales de l'évènement 
qui diffèrent de seulement 1.4 secondes par rapport à une durée observée de plus de 6h \citep{Le-V1845e}.

\section{La comète Lexell}

Le 22 novembre 1843, Hervé Faye découvre une nouvelle comète à l'Observatoire de Paris. Cette découverte est annoncée 
dès la séance du 27 novembre de l'Académie \citep{Faye1843a} avec les observations des 22 et 24 novembre. Elle suscite 
immédiatement l'intérêt. La comète est en un premier temps considérée comme parabolique,
avec des premières estimations de ses éléments \citep{Faye1843a}.

\bqs
J'ajouterai que M. Faye n'a pas réussi jusqu'ici à représenter 
convenablement les observations par une orbite parabolique. Ce jeune
astronome attend que l'état du ciel lui ait permis d'obtenir une nouvelle position
de l'astre qu'il a découvert, pour entreprendre la détermination des
éléments elliptiques \citep{Cras1844a}.
\eq

Mais en janvier 1844, avec l'augmentation du nombre des observations, 
il semble bien que l'orbite de la comète ne soit pas parabolique, mais elliptique. H. Goldschmidt, un élève de Gauss, 
fournit des éléments de l'orbite  de la comète, qui s'accordent aux observations à moins de 2' des observations
\citep{Cras1844b}. Dans le compte rendu de la séance du 15 janvier,  apparait une  {\it note rajoutée le mercredi},
donc après la séance de l'Académie du lundi, effectuant le rapprochement avec la comète Lexell, observée en 1770.

\bqs
M. Faye a remarqué à ce sujet que, d'après les éléments ci-dessus, le nouvel
astre a dû passer, vers son aphélie, assez près de Jupiter pour en éprouver des perturbations
sensibles. On pourrait donc supposer qu'il présente un cas analogue à celui de la
comète de Lexell, dont l'orbite parabolique fut transformée par l'attraction de Jupiter en une
orbite elliptique, et redevint plus tard parabolique par l'action perturbatrice de la même
planète.
\eq

Deux semaines plus tard, les éléments de l'orbite calculés par Faye sont publiés dans les CR, et précisent l'orbite initiale 
de Goldschmidt dans une note élaborée par les {\it Commissaires} Arago, Mathieu, Damoiseau, Liouville, et Mauvais  \citep{Cras1844c}.
Ils seront suivi trois semaines plus tard de la nouvelle détermination de Plantamour, elliptique  cette fois-ci, et en accord avec les précédentes 
\citep{Cras1844d}.  Finalement,  les CR de la séance du 25 mars 1844 publient  la version révisée des éléments elliptiques obtenus 
par H. Goldschmidt \citep{Cras1844e}. Ce dernier retrace le parcours de la comète et s'interroge sur ses rencontres
proches avec Jupiter 

\bqs
L'orbite correspondant à mes troisièmes éléments s'approche extrêmement
de l'orbite de Jupiter, à 210 degrés de longitude. La plus petite distance
de ces deux orbites s'élève ici à 0.1199 (la moyenne distance entre la Terre
et le Soleil étant prise pour unité). La comète était à cet endroit, pour la dernière
fois, le 23 décembre 1838; mais alors ... sa distance à la comète s'élevait à 2,254, et selon des calculs dans
lesquels, il est vrai, je n'ai pas eu égard aux perturbations très considérables
que la comète éprouve de la part de Jupiter, les deux corps ne se trouvèrent
jamais simultanément très proches de cet endroit pendant les dix dernières
révolutions de la comète.
\eq

On ne peut qu'être impressionné par la  rapidité avec 
laquelle le débat scientifique s'exprime. Rapidité, qui n'exclut pas une validation des 
informations, grâce au système de rapporteurs (les {\it Commissaires}) mis en place au sein de l'Académie des Sciences. 
Il n'y a pas, à ma connaissance, de système de publication équivalent à l'heure actuelle. Mais le feuilleton 
sur la comète continue et Arago fait lecture, dans la séance du 22 avril, d'une lettre 
de M. Valz suggérant à nouveau que la comète découverte par Faye serait celle de 1770

\bqs
Je viens vous faire part du résultat extraordinaire auquel je suis parvenu; 
c'est que la dernière comète ne serait autre que celle de 1770, que
Jupiter nous avait enlevée en 1779, et qu'il nous rendrait de nouveau, ainsi qu'il était déjà arrivé en 1767. 
Cela est sans doute fort extraordinaire, mais 
n'en est pas moins dans l'ordre  des possibilités et même des probabilités; \citep{Cras1844f}
\eq

\dots

\bqs
D'après ce qui précède, Jupiter, ce dominateur, ce tyran pourrait-on
dire de notre système planétaire, semble destiné à jouer un rôle fort important
dans la transformation des orbites cométaires, ainsi qu'on peut en juger
par la comète de 1770 \citep{Cras1844f}
\eq

A la fin de la lecture de cette lettre de Valz, Cauchy annonce que Le Verrier, lui aussi 
{\it s'est livré, sur la comète de 1770, à des recherches toutes semblables}. 
Dans la séance suivante, la lettre de Le Verrier à Cauchy montre qu'il occupe bien le terrain 
\citep{Le-V1844a}, et il attaque avec vigueur

\bqs
Je vous suis très reconnaissant d'avoir bien voulu, à l'occasion de la Lettre
de M. Valz, annoncer immédiatement que je vous avais communiqué, peu
de temps auparavant, des résultats semblables à ceux qu'envoie aujourd'hui
cet astronome. Vos remarques me permettront de continuer mon travail, et
de le présenter plus tard à l'Académie, sans qu'on puisse m'accuser de m'emparer
des idées d'autrui.

Les mêmes considérations qui ont guidé M. Valz m'ont porté à penser
que la comète de M. Faye était la même que celle de Lexell; mais je n'ai
pas cru devoir entretenir l'Académie de résultats aussi vagues avant d'être
parvenu à leur donner quelque précision. Il m'a paru inutile de soulever
cette question sans apporter en même temps une solution satisfaisante, à laquelle
je travaille.
\eq

Valz revient à la charge dans une nouvelle lettre à Arago \citep{Cras1844g}, 
tandis que des premiers extraits de la {\it Théorie de la comète périodique de 1770}
de Le Verrier sont présentés dans la séance du 11 novembre 1844 \citep{Le-V1844d}. 

\bqs
Le travail que je présente aujourd'hui à l'Académie, et qui est un fragment
étendu des recherches que j'ai entreprises sur les comètes, peut se diviser
en six sections.
\eq
 
Dans sa section 4, Le Verrier  constate la dégénérescence du problème. 

\bqs
Lorsqu'en éliminant entre des équations du premier degré, on finit par
tomber sur une équation finale dont tous les termes se détruisent, on en
conclut que le système est indéterminé. Or, le cas dans lequel nous nous
trouvons ici approche de ce cas extrême de l'indétermination; et nous devons
dire que les quatre mois d'observations sont insuffisants pour déterminer
d'une manière précise tous les éléments de l'orbite.
\eq

Il a alors l'idée de ne pas chercher une seule solution, mais une famille 
continue de solutions, dépendant d'un seul paramètre. 

\bqs
Cette indétermination m'avait jeté dans un grand embarras; car je savais
que l'action de Jupiter sur la comète serait fort différente, suivant que je
m'arrêterais à l'une ou à l'autre des solutions auxquelles je voyais qu'on pouvait
également arriver. Cependant, en considérant mes différents systèmes
de solutions, et en les rapprochant de la solution donnée par Lexell, je reconnus
que les variations que subissaient les éléments, quand on passait de
l'un à l'autre de ces systèmes, suivaient une marche progressive. Je fus ainsi
conduit à penser que si les valeurs absolues des différents éléments étaient
mal déterminées, on pourrait au contraire les considérer comme des fonctions
bien définies d'une même arbitraire \citep{Le-V1844d};
\eq

Le Verrier se rend compte de la très grande dépendance de la solution par rapport aux conditions initiales. 
Comme le faisait remarquer Giovanni Valsecchi 
lors du dernier congrès de mécanique céleste \citep{Vals2017a},  
 il est véritablement en train d'étudier le comportement d'un système chaotique. 

\bqs
 En sorte qu'il existe un système
d'éléments satisfaisant aux observations de 1770, avec la même précision
que celui donné par la méthode des moindres carrés, et pour lequel il se
pourrait, à tout prendre, que la comète fût allée heurter Jupiter! \citep{Le-V1844d}.
\eq

Après avoir mis au point sa méthode générale, Le Verrier l'applique dans  la détermination 
des éléments de la comète de Faye, qu'il fait donc dépendre d'un paramètre $\mu''$, afin de 
tenir compte de la dégénérescence de cette détermination \citep{Le-V1845c}.

La méthode mise au point par Le Verrier aura des suites. C'est en effet 
celle-ci qui est le plus souvent utilisée à l'heure actuelle pour expliciter les orbites 
des comètes ou des astéro\"ides présentant des rencontres proches avec les planètes, 
même si les utilisateurs actuels en ont souvent oublié l'origine  \citep{ValsMila2003a,Vals2007a}.
La famille continue d'orbites déterminée par le Verrier est maintenant appelée  {\it Line of Variation} (LoV), 
dont l'article fondateur reste celui de Le Verrier du CR du 11 novembre 1844 \citep{Le-V1844d}.
Le Verrier reviendra sur le mouvement de la comète Lexell ultérieurement, en 1848
pour réaffirmer que sa méthode lui permet de mieux identifier les comètes à l'issue d'une rencontre proche avec Jupiter, 
et en particulier 

\bqs
Toutefois, l'ensemble des éléments de l'orbite de la comète, postérieurement
à 1779, est loin d'être arbitraire. Chacun de ces éléments se
trouve encore fonction de l'indéterminée $\mu$. 
\dots C'est ainsi que je suis parvenu à démontrer que la comète
périodique de 1770 n'a rien de commun avec les comètes périodiques découvertes
en 1843 et 1844 par MM. Faye et de Vico \citep{Le-V1848d}.
\eq

Il est effectivement admis aujourd'hui que les comètes Faye et Lexell sont différentes, cette dernière ayant été perdue de vue 
après 1779 \citep{ValsMila2003a}.

Le Verrier ne poursuivra pas immédiatement ces travaux, car il va être absorbé par une tâche plus 
importante, la recherche d'une nouvelle planète dans  le système solaire. Il reviendra plus tard 
sur l'analyse de la trajectoire de la comète Faye dans les CR du 9 décembre  1850 \citep{Le-V1850b}, 
après que la comète ait été à nouveau observée, le 28 novembre de la même année. 
Le Verrier utilisera alors sa méthode, avec l'introduction du paramètre $\mu$, pour préciser les éphémérides de la 
comète  \citep{Le-V1850b,Le-V1850f}.

\section{ L'entrée à l'Académie des Sciences} 

Un attrait des Comptes Rendus, est qu'ils rendent compte de l'ensemble des débats de l'Académie des Sciences, 
y compris les résultats des votes des comités secrets pour l'élection des nouveaux membres.
On y apprend par exemple que dans la séance du 3 juin 1839,  
Gustave de Pontécoulant  a échoué à l'élection de membre de l'Académie dans la section d'astronomie, avec 
18 voix contre 29 pour Joseph Liouville.   
En revanche, dans la séance du 17 février 1840, Jacques Babinet est élu dans la section de physique 
avec 39 voix ( contre 19 à Despretz, et 2 à Péclet). On discutera plus loin de ses démêlés avec Le Verrier.

Lors de la séance du 20 novembre 1843, Le Verrier échoue en seconde position avec 7 voix contre 
Félix Mauvais qui obtient 30 voix et est donc élu sans ambiguité (Bravais, 6 voix, Largeteau, 6 voix, Bouvard, 1 voix).
En revanche, 3 ans plus tard, le 26 janvier 1846, Le Verrier obtient 44 voix, contre 9 pour Bouvard. Il est donc largement 
élu, ce qui est normal, au vu des travaux impressionnants qu'il a  déjà effectués à cette date.  

\section{Premières études des perturbations d'Uranus. Nouvelles polémiques.} 

Le premier travail de Le Verrier sur Uranus est publié dans le CR  du 28 mars 1842. 
C'est l'occasion pour lui de croiser le fer avec Charles Delaunay qui propose le 14 mars 1842 une correction 
dans la solution du mouvement d'Uranus \citep{Dela1842a}.

\bqs
M. Delaunay annonce un terme de cette espèce
dans la longitude d'Uranus. Il dépendrait de l'argument $4n^{VI}-n^V$, l'angle
$3n^{VI}-n^V$, étant fort petit. On peut, ce me semble, démontrer simplement
qu'un pareil terme n'existe pas réellement \citep{Le-V1842a}.
\eq

Delaunay rétorque dans une communication du 18 avril 1842, tout en reconnaissant 
que le terme en question est déjà pris en compte. 

\bqs
M. Le Verrier a adressé à l'Académie, dans sa séance du 28 mars, uue
Note dans laquelle il cherche à prouver que  une des deux inégalités que
j'avais annoncées précédemment n'existe pas réellement: je me propose de
faire voir aujourd'hui que M. Le Verrier s'est trompé en attribuant à ces
inégalités une cause tout autre que celle qui les produit.

\dots 

Il résulte donc de ce qui précède, que les inégalités que j'avais annoncées
existent bien; qu'elles ne sont pas données, dans la Mécanique
céleste, au chapitre de la théorie d'Uranus; et que, si l'on avait poussé
plus loin les approximations, on les aurait trouvées. Ce sont ces raisons
qui me les ont fait regarder comme nouvelles. J'ajouterai cependant que
j'ai reconnu depuis que, si ces inégalités ne sont pas données explicitement
dans la Mécanique céleste, elles y sont implicitement comprises, et ont
été, comme telles, employées dans la construction des tables \citep{Dela1842b}.
\eq

Le Verrier précisera ces divergences dans le CR du 3 mai. Le problème de 
la compréhension des écarts entre la théorie et les observations d'Uranus est déjà présent,
et il est essentiel pour Le Verrier de ne pas brouiller les pistes. 
Il n'hésite pas à rétorquer vivement à Delaunay en mettant en évidence le fait que les termes 
qu'il introduit sont le résultat d'une différence de méthode dans le développement
des termes perturbatifs. 
Pour Le Verrier, en suivant { \it La Mécanique Céleste} de Laplace, qui est l'ouvrage de référence, 
il faut d'abord calculer les modifications du moyen mouvement, et ensuite l'introduire 
dans les solutions, alors que Delaunay utilise le moyen mouvement non perturbé, 
et exprime les différences par de nouveaux termes de perturbation. 
Ces premières escarmouches ne se tariront pas, et Le Verrier et Delaunay resteront en conflit 
jusqu'à la mort de ce dernier en 1872 par noyade accidentelle dans la baie de Cherbourg.

\bqs
Cette communication entraînait avec elle de graves conséquences. S'il
était constaté qu'on avait négligé dans la théorie d'Uranus plusieurs termes,
tels que ceux qui étaient indiqués, on pouvait espérer d'avoir la clef des
grands écarts des tables de cette planète. Si au contraire on y introduisait à
tort de nouveaux termes, on obscurcissait pour longtemps une théorie
déjà si peu claire. Aussi, quand je me fus convaincu que les perturbations
annoncées ne devaient pas être ajoutées aux tables existantes, il me sembla
que l'intérêt de la science exigeait que je le fisse connaître.

\dots 

Cependant, en même temps que M. Delaunay convient, dans la dernière
partie de sa note, de la justesse de mes observations, il s'efforce, dans
la première partie, de prouver que j'ai complètement tort. D'après lui,
je ne serais parvenu à une conclusion exacte que par une suite de faux
raisonnements. Je crains que l'auteur, entraîné par la vivacité de sa critique,
n'ait pas aperçu qu'elle reposait sur des erreurs presque matérielles \citep{Le-V1842b}.
\eq

Il n'y a pas de doute, Le Verrier tient à  faire comprendre qu'il est à son époque 
un acteur incontournable 
de la mécanique céleste. 

\section{Les perturbations d'Uranus et la découverte de Neptune} 
En 1845, Le Verrier est même le  maître incontesté de la mécanique céleste. 
Sa théorie de Mercure est le nouveau standard des solutions de perturbations planétaires, et 
ses travaux sur les comètes et les petites planètes de grande excentricité  font preuve d'une 
grande originalité, et restent des références utilisées actuellement. 
Il sera élu  sans ambigu\"ité à l'Académie en janvier 1846. 
Comme il a été dit plus haut, son principal sujet d'étude était alors le problème de la 
détermination des orbites de comètes et en particulier de la comète Lexell, mais sur 
l'avis d'Arago, il abandonnera temporairement la comète Lexell pour se consacrer 
aux perturbations d'Uranus.

\bqs
\dots chaque jour Uranus s'écarte de plus en plus de la route qui lui est tracée dans les \'Ephémérides. 

Cette discordance préoccupe vivement les astronomes, qui ne sont pas habitués à de pareils mécomptes. 
Déjà elle a donné lieu à un grand nombre 
d'hypothèses. On est même allé jusqu'à mettre en doute que le mouvement d'Uranus fût rigoureusement 
soumis au grand principe de la gravitation universelle. 
 
Dans le courant de l'été dernier, M. Arago voulut bien me représenter 
que l'importance de cette question imposait à chaque
astronome le devoir de concourir, autant qu'il était en lui, à en éclaircir quelque point. J'abandonnai
donc momentanément, pour m'occuper d'Uranus, les recherches que j'avais entreprises sur les comètes, \dots
\citep{Le-V1845f}.
\eq

Comme dans ses travaux précédents, Le Verrier ne se contente pas d'une adaptation des solutions existantes 
pour Uranus. Il veut reprendre à la base l'ensemble du travail. Sa première étape consiste donc dans 
l'élaboration d'une nouvelle solution complète pour le mouvement d'Uranus. Ses premières conclusions 
sont présentées à l'Académie le 10 novembre 1845, soit seulement deux à trois mois après que Le Verrier eut commencé son travail.

\bqs
Le Mémoire actuel a pour but d'établir la forme et la grandeur des termes que
les actions perturbatrices de Jupiter et de Saturne introduisent dans les expressions 
des coordonnées héliocentriques d'Uranus. Les formules, ainsi obtenues
seront comparées aux observations de Paris et de Greenwich dans  une seconde communication \citep{Le-V1845f}.
\eq

Ce qui frappe dans ce travail, est la rapidité avec 
laquelle l'étude est conduite. On entend  souvent  dire que Le Verrier avait à sa disposition une 
armée de calculateurs. Il est juste de dire ici que cette rumeur est totalement fausse
pour ce qui concerne la découverte de Neptune. Il n'aura l'aide de calculateurs qu'une fois à la direction de 
l'Observatoire de Paris, en 1854. D'ailleurs, l'intégralité des manuscrits de Le Verrier pour ses 
calculs de la découverte de Neptune a été conservée à l'Observatoire de Paris, et comprend 
environ 1300 pages manuscrites, apparemment toutes écrites de la main de Le  Verrier. 

Le Verrier présente ses {\it Recherches sur le mouvement d'Uranus} à l'Académie dans la séance du 1er juin 1846 \citep{Le-V1846d}.
Après la découverte d'Uranus en 1781, les astronomes ont retrouvé des observations anciennes de la planète 
qui avaient été consignées, de 1690 à 1771, sans en reconnaitre la nature planétaire. 
C'est avec ces observations, additionnées des observations plus récentes obtenues depuis sa découverte que Bouvard 
avait établi les tables d'Uranus, sans arriver à concilier les observations anciennes avec les modernes.  
En tenant compte de manière plus complète des perturbations de Jupiter et Saturne, Le Verrier 
pensait en un premier temps résoudre ces problèmes, mais 

\bqs
En tenant compte, dans ce but, des altérations que les
perturbations négligées avaient dû produire dans les éléments de l'ellipse,
je reconnus que, si l'écart des Tables, en 1845, était effectivement notablement
 diminué  par l'emploi des nouvelles expressions, il restait encore très considérable
 et supérieur aux erreurs des observations \citep{Le-V1846d}. 
\eq 

Ne faisant pas, à juste titre, entièrement confiance aux travaux anciens, Le Verrier entreprend aussi 
une nouvelle réduction complète de l'ensemble des observations.

\bqs
L'importance du sujet me faisait une loi de tout revoir, de tout vérifier
moi-même. A l'égard des anciennes observations, j'ai réduit de nouveau
celles de Flamsteed, Bradley, Mayer et Lemonnier; et, parmi les nouvelles,
j'en ai choisi deux cent soixante-deux, faites principalement aux instants des
oppositions et des quadratures \citep{Le-V1846d}.
\eq

Le Verrier examine toutes les hypothèses avancées pour expliquer l'écart observé entre le calcul et les 
observations. Il ne veut envisager un changement possible des lois de gravitation 
qu'en dernier recours, après avoir examiné toutes les autres hypothèses 

\bqs
Je ne m'arrêterai pas à cette idée que les lois de la gravitation pourraient
cesser d'être rigoureuses, à la grande distance à laquelle Uranus est situé du
Soleil. Ce n'est pas la première fois que, pour expliquer des inégalités dont
on n'avait pu se rendre compte, on s'en est pris au principe de la gravitation
universelle. Mais on sait aussi que ces hypothèses ont toujours été anéanties
par un examen plus approfondi des faits. L'altération des lois de la gravitation
serait une dernière ressource à laquelle il ne pourrait être permis
d'avoir recours qu'après avoir épuisé l'examen des autres causes, qu'après les
avoir reconnues impuissantes à produire les effets observés \citep{Le-V1846d}.
\eq

Il réfute aussi les explications possibles par la résistance de l'éther, par l'existence d'un gros satellite, ou par une comète 
qui perturberait le mouvement de la planète. 

\bqs
Il ne nous reste ainsi d'autre hypothèse à essayer que celle d'un corps
agissant d'une manière continue sur Uranus changeant son mouvement
d'une manière très lente  \citep{Le-V1846d}.
\eq

Des considérations simples lui font dire que cette planète doit se trouver bien au-delà d'Uranus. 
De là, il utilise la loi empirique de Bode de répartition des planètes pour 
prendre comme première hypothèse que la planète se trouverait à une distance du soleil double de celle d'Uranus, 
dans le plan de l'écliptique et pose la question 

\bqs 
Est-il possible que les inégalités d'Uranus soient dues à l'action d'une
planète, située dans l'écliptique, à une distance moyenne double de celle
d'Uranus? Et, s'il en est ainsi, où est actuellement située cette planète?
Quelle est sa masse ? Quels sont les éléments de l'orbite qu'elle parcourt \citep{Le-V1846d} ? 
\eq
 
Il fournit alors une première approche pour la longitude héliocentrique de la planète, 
donnée par 
$$
\nu = 314^\circ,5 + 12^\circ,25 \alpha + {1\over m} \left\{ 20^\circ,82 - 10^\circ, 79\alpha -1^\circ ,14 \alpha^2\right\}
$$
où $m$ est la masse de la planète recherchée, et $\alpha$ un paramètre. 
Mais Le Verrier précise bien 

\bqs
 Le travail dont
je viens de présenter un extrait à l'Académie doit être considéré comme une
ébauche d'une théorie qui commence \citep{Le-V1846d}.
\eq

Deux mois plus tard, dans la séance du 31 août 1846 de l'Académie, Le Verrier présente les nouvelles
conclusions de son travail qui a beaucoup avancé. Il est alors en mesure de donner les éléments elliptiques complets de 
la planète, supposée dans le plan de l'écliptique \citep{Le-V1846e}

\begin{center}
\small
\begin{tabular}{ll}
   Demi-grand axe &  36.154 UA \\
 Période sidérale &  217.387 an  \\
  Excentricité &  0.10761 \\
Longitude du périhélie & $284^\circ$ 45' \\
Longitude moyenne au 1er janvier 1847 \phantom{00000}& $318^\circ$ 47' \\
Masse   & 1/9300 masse solaire \\ 
\end{tabular}
\end{center}

Et pour la position actuelle de la planète au 1er janvier 1847
 
\begin{center}
\small
\begin{tabular}{ll}
Longitude héliocentrique vraie \phantom{000000000000} & $326^\circ$ 32' \\
Distance au Soleil  & 33.06 UA  \\ 
\end{tabular}
\end{center}
 
Il cherche aussi à inciter les observateurs à rechercher la planète, en donnant des détails pratiques  sur ses conditions 
d'observation 

\bqs
L'opposition de la planète a eu lieu le 19 août dernier. Nous sommes
donc actuellement à une époque très favorable pour la découvrir. L'avantage
qui résulte de sa grande distance angulaire au Soleil ira en diminuant
sans cesse; mais, comme la longueur des jours décroît maintenant très rapidement
dans nos climats, nous nous trouverons longtemps encore dans une
situation favorable aux recherches physiques qu'on voudra tenter \citep{Le-V1846e}.

\dots 

Nous trouverons ainsi, qu'au moment de l'opposition, la nouvelle
planète devra être aperçue sous un angle de 3'',3. Ce diamètre est tout à fait
de nature à être distingué, dans les bonnes lunettes \citep{Le-V1846e}.
\eq

Le Verrier s'autorise aussi à donner des limites pour ses déterminations. Limites pour lesquelles il s'avance un peu 
et qui lui seront reprochées plus tard. 

\bqs
Le demi-grand axe de l'orbite, auquel j'ai trouvé pour valeur la plus
précise 36,154, ne peut varier qu'entre les limites 35,04 et 37,90. Les durées
extrêmes correspondantes de la révolution sidérale sont 207 et 233 ans
environ \citep{Le-V1846e}.
\eq

Il montre par ailleurs, qu'une fois cette nouvelle planète prise en compte, les résidus des observations 
se réduisent considérablement à quelques secondes d'arc, aussi bien   pour les observations récentes
que les anciennes.  Ceci l'encourage à écrire 

\bqs
Je terminerai cette analyse, par une remarque qui me paraît très propre
à porter dans les esprits la conviction que la théorie que je viens d'exposer
est l'expression de la vérité  \citep{Le-V1846e}.
\eq 


Les derniers résultats sont présentés à l'Académie immédiatement après la découverte, 
 au cours de la séance du 5 octobre 1846 \citep{Le-V1846f}. 
 Le Verrier commence par présenter les calculs qu'il a conduits pour la détermination de la latitude 
de l'objet recherché, ce qui correspond à la détermination de l'inclinaison du plan de son orbite. Il 
trouve une inclinaison minimale de $4^\circ 38' $  avec le plan de l'orbite d'Uranus. 
Ce n'est qu'après qu'il rend compte des circonstances de l'observation de la planète par Johann Gottfried  Galle, 
à Berlin,  dans la nuit du 23 septembre, le soir même du jour où il reçut de Le Verrier la prédiction de sa 
position calculée. 

\bqs
J'avais
écrit le 18 septembre à M. Galle, pour réclamer son bienveillant concours;
cet habile astronome a vu la planète le jour même où il a reçu ma Lettre \citep{Le-V1846f}.
\eq

La longitude héliocentrique observée est de $327^\circ 24'$, celle prédite par Le Verrier, et donnée 
dans les CR du 31 août 1846 vaut $326^\circ 32'$ \citep{Le-V1846e}, soit une différence de seulement $52'$, ce qui fait dire 
à Le Verrier qui n'en espérait sans doute  pas tant  

\bqs 
Ainsi la position avait été prévue à moins d'un degré près. On trouvera
cette erreur bien faible, si l'on réfléchit à la petitesse des perturbations dont
on avait conclu le lieu de l'astre \citep{Le-V1846f}.
\eq

L'ensemble des recherches de Le Verrier ayant conduit à la découverte de Neptune 
est publié sous la forme d'un volume séparé \citep{Le-V1846l}
qui comprend  toutes les étapes de ses calculs, mais pour mieux comprendre la démarche de celui-ci, nous avons entrepris cette année, 
dans le cadre d'un travail de thèse, la lecture de la totalité des notes manuscrites de Le Verrier, 
avec pour but de reproduire l'intégralité de ses calculs ayant conduit à la découverte de Neptune. 
Ce travail n'est encore qu'à ses débuts mais  nous permet déjà d'être admiratif devant l'ampleur du travail  fourni par Le Verrier 
pour la découverte de Neptune. 

A la suite de l'intervention de Le Verrier, Arago lit la lettre que  Galle a  envoyée à Le Verrier le 25 septembre 1846,
au lendemain de la découverte. 

\bqs
La planète dont vous avez signalé la position existe réellement. Le jour
même où j'ai reçu votre Lettre, je trouvai une étoile de huitième grandeur,
qui n'était pas inscrite dans l'excellente carte Hora XXI (dessinée par M. le
docteur Bremiker), de la collection de Cartes célestes publiée par l'Académie
royale de Berlin. L'observation du jour suivant décida que c'était la
planète cherchée \citep{CRAS1846a}.
\eq 

Arago poursuit par l'éloge de Le Verrier 

\bqs
M. Le Verrier a aperçu le nouvel astre sans avoir
besoin de jeter un seul regard vers le ciel; il l'a vu au bout de sa plume \citep{CRAS1846a}.
\eq 

Puis suivent les félicitations provenant de diverses parties du monde, et en tout premier lieu de  Encke, 
alors directeur de l'Observatoire de Berlin, qui avait autorisé Galle à effectuer cette recherche. 

\bqs
Votre nom
sera à jamais lié à la plus éclatante preuve de la justesse de l'attraction
universelle qu'on puisse imaginer \citep{CRAS1846a}.
\eq 

Au cours de  cette même séance, se pose la question du nom de la nouvelle planète. Arago, alors très ami
avec Le Verrier, propose de donner le nom de celui-ci à la planète, selon la coutume qui prévaut pour la découverte de 
comètes. 

\bqs
Je prends l'engagement, a dit en terminant M. Arago, de ne jamais appeler la nouvelle planète, 
que du nom de Planète de Le Verrier. Je croirai
donner ainsi une preuve irrécusable de mon amour des sciences, et suivre
les inspirations d'un légitime sentiment de nationalité \citep{CRAS1846a}.
\eq  

\section{Polémiques sur la nouvelle  planète}
Cependant, dès la séance du 19 octobre 1846, des voix dissonantes se font entendre au sein du concert d'éloges en 
faveur de Le Verrier. Arago prend le devant pour répondre aux critiques \citep{Arag1846a}.

\bqs
La planète Le Verrier est-elle la première dont on ait prévu l'existence et la position
par la théorie ?
\eq

Arago montre bien que les cas précédents sont plus dus au hasard qu'à une prédiction précise, 
 et que dire, en se basant sur la loi empirique de Titius-Bode, qu'il y avait {\it 
probablement une nouvelle planète au delà d'Uranus} \dots {\it n'a aucun fondement théorique,
qu'elle manque d'exactitude, qu'elle ne pouvait pas indiquer la direction
dans laquelle il fallait placer le nouvel astre, ni même servir à constater
son existence.} \citep{Arag1846a}

\bqs
La planète avait-elle été vue avant que M. Le Verrier annonçât son existence et en fixât la
position ?
\eq

Il n'est pas impossible que Neptune ait été observée auparavant, comme cela a été le cas pour Uranus, sans que les observateurs 
aient pu en discerner la nature planétaire. Mais Arago montre que les observations de Cacciatore sont diamètralement opposées à 
la position de la nouvelle planète. Quant à celles de Wartmann, elles en diffèrent de plus de 17 degrés \citep{Arag1846a}.

\bqs
Les calculs de M. Le Verrier avaient-ils assigné à la nouvelle planète une position aussi
voisine de la véritable, que l'ont proclamé dès l'abord, avec admiration, les astronomes
allemands ?
\eq

Sur cette question, suscitée par un texte de Giovanni Plana, Arago montre que les calculs de ce dernier sont 
entachés d'erreur, et qu'il a dû considérer des valeurs approchées des positions, données sans doute dans le but 
d'en simplifier l'expression pour le grand public. 
Plus importante est la question de priorité qui apparaît une fois la planète découverte. 

\bqs
De la question de priorité, soulevée, il y a quelques jours, en Angleterre, par sir John Herschel,
M. Airy, directeur de l'Observatoire de Greenwich, et M. Challis, directeur de l'Observatoire
de Cambridge.
\eq 

Dans une lettre du 1er octobre 1846, John Herschel,  le fils de William Herschel, le découvreur d'Uranus, 
fait état du travail de John Couch Adams, qui s'est livré, à Cambridge, de manière indépendante, 
à un calcul similaire à celui de Le Verrier. Adams avait remis ses calculs à l'Astronome Royal George Airy, 
qui ne leur avait pas adressé toute l'attention qu'ils méritaient \citep[voir][]{Lequ2009a}.
Après la découverte de Neptune, Airy se rend compte de son erreur, et envoie une lettre à Le Verrier le 
14 octobre 1846 dont Arago va donner lecture en séance \citep{Arag1846a}.

\bqs
Je ne sais si vous êtes instruit que des recherches collatérales, faites en Angleterre, 
avaient conduit précisément au résultat obtenu par vous.
Probablement je serai appelé à m'expliquer sur ces recherches. Si, dans
ce cas, j'accorde des éloges \dots à d'autres, je désire que vous
ne les considériez pas comme affaiblissant en aucune manière l'opinion
que j'ai sur vos droits. Vous devez, sans aucun doute, être considéré
comme celui qui a réellement prédit la position de la planète. Je puis
ajouter que les investigations anglaises n'étaient pas, je crois \dots,
tout à fait aussi étendues que celles dont on vous est redevable. Je les
connaissais d'ailleurs avant d'être informé des vôtres.
\eq 

Après lecture de cette lettre, pour le moins alambiquée, Arago poursuit \citep{Arag1846a}.

\bqs
Ces deux pièces,\dots , ne peuvent être considérées que
comme des escarmouches. Nous serons en pleine bataille en donnant la
traduction suivante d'une Lettre de M. Challis, directeur de l'Observatoire
de Cambridge, adressée au journal anglais l'Athenaeum.
\eq

Dans sa lettre, Challis fait état des travaux de Adams

\bqs
En septembre 1845, M. Adams me communiqua
les valeurs qu'il avait obtenues pour la longitude héliocentrique, 
l'excentricité de l'orbite, la longitude du périhélie, et la masse d'une planète
extérieure présumée, valeurs qu'il avait entièrement déduites des irrégularités d'Uranus, 
non représentées par les perturbations. En octobre, 
il communiqua les mêmes résultats, un peu corrigés, à l'astronome royal.
\eq

Il mentionne aussi les recherches pour observer la planète, après qu'il eut reçu les prédictions 
de Le Verrier. En voyant que ces prédictions étaient similaires à celles d'Adams, il entreprit 
de balayer une large portion du ciel

\bqs
Mes observations suivantes eurent lieu le 12 du mois d'août; 
je rencontrai ce jour-là une étoile de huitième grandeur, dans la zone que 
j'avais explorée le 3o juillet, et qui alors ne contenait pas cette étoile.
Conséquemment, celle-ci était la planète.
\eq

Mais Challis ne la reconnait pas car il n'a effectué ces comparaisons qu'une fois annoncée la découverte 
de la planète. Cependant, il revendique la possibilité de nommer la planète
en raison de ces travaux, et propose {\it Oceanus} à quoi Arago rétorque 
vertement, en faisant état du manque de publications de la part de Adams \citep{Arag1846a}.

\bqs
M. Arago a remarqué qu'il n'y est fait mention d'aucune publication
du travail de M. Adams, ni de rien qui en ait même l'apparence. Cette
circonstance, a dit le secrétaire, suffit pour mettre fin au débat. Il n'existe
qu'une manière rationnelle et juste d'écrire l'histoire des sciences : c'est de
s'appuyer exclusivement sur des publications ayant date certaine; hors de
là, tout est confusion et obscurité. M. Adams n'a pas imprimé, même aujourd'hui,
une seule ligne de ses recherches; il ne les a communiquées à aucune
Société savante : M. Adams n'a donc pas le moindre titre valable pour figurer
dans l'histoire de la découverte de la nouvelle planète.
\eq

Les CR sont particulièrement précieux sur ce point, car ils retracent en détail le débat
virulent qui eut lieu à l'Académie des sciences, 
qui ne serait autrement accessible qu'à travers les quelques lettres entre les intéressés, et 
les articles de journaux, forcément beaucoup plus partiaux. 
Après avoir démonté les arguments de Herschel, de Airy et de Challis, Arago conclut \citep{Arag1846a}.

\bqs
M. Adams n'a le droit de figurer, dans l'histoire de la découverte de la
planète Le Verrier, ni par une citation détaillée, ni même par la plus légère
allusion.
\eq

Peut-être inspiré par ces vives attaques, Joseph Bertrand, jeune mathématicien prodige d'alors 24 ans 
soumet dans la séance du 26 octobre 1846 un mémoire dénonçant vigoureusement les 
{\it erreurs  tellement graves} des Mémoires mathématiques de Challis \citep{Bert1846a}. 
De son côté, Galle avait proposé le nom de {\it Janus}, ce qui n'était pas du goût de Le Verrier
qui le fit savoir dans le  {\it Le Journal des Débats} du 30 septembre 1846. Ce quotidien, 
qui rendait compte régulièrement des débats de l'Académie des sciences, 
précise que \citep{jdb1846a}

\bqs
 M. Le Verrier, à qui revient évidemment le droit de nommer le nouvel astre,
n'accepte pas le nom trop significatif de Janus. Il
donne au reste son adhésion à toute autre désignation, telle que Neptune, par exemple, qui aura l'assentiment des 
astronomes.
\eq

Cette proposition de nommer la planète {Neptune} sera reprise par le Bureau des longitudes 
le jour même sans prendre de résolution très claire \citep{bdl1846a}

\bqs 
On discute les propositions qui ont déjà été faites touchant le nom à donner à la planète de Mr. Le Verrier 
et, en particulier le nom de Neptune.
\eq

Ce nom de  {\it Neptune} sera largement acceptée par la communauté, 
contrairement à l'avis ultérieur de Le Verrier qui, finalement, aurait sans doute préféré voir adoptée la proposition d'Arago 
d'appeler la planète de son nom \citep{Koll2009a,Lequ2009a}. Il sera définitivement adopté par le Bureau 
des longitudes dans sa séance du 28 juillet 1847 \citep{bdl1847a}. 

\bqs
Le Bureau des longitudes n'avait pris jusqu'à présent aucune décision relativement au choix du nom qu'il 
convenait de donner à la nouvelle planète. Celui de Neptune ayant aujourd'hui prévalu parmi les astronomes, 
le Bureau se décide à l'adopter. M. Arago qui avait demandé que la nouvelle planète portât le nom de 
M. Leverrier a cru devoir s'abstenir de prendre part à cette résolution.
\eq 

Il faut noter que Le Verrier, qui avait été élu 
membre adjoint du Bureau des longitudes  à l'unanimité le 14 octobre 1846 \citep{bdl1846b}, n'était pas présent 
à cette séance du Bureau du 28 juillet, car il en avait démissionné en février 1847 \citep{Lequ2009a,bdl1847b}. 
A la demande du Ministre de l'Instruction Publique, il reviendra sur cette démission en août 1847 \citep{bdl1847c}.

\section{Nouvelles attaques}

Avec une meilleure détermination des éléments de la nouvelle planète, 
il apparaît des différences notables entre les prédictions de Le Verrier, et les éléments observés.
Binet effectue une première estimation de la distance de la planète au Soleil, en prenant 
l'hypothèse simplificatrice d'une orbite circulaire, et trouve $30.245$ UA \citep{Bine1846a}, valeur sensiblement moindre que
celle proposée  par Le Verrier de 33.06 UA \citep{Le-V1846e}.
Par ailleurs, la découverte d'un satellite de Neptune,  Triton, par Lassell le 10 octobre 1846
\citep{Lass1846a} permet une estimation précise de la masse de Neptune bien moindre que celle que Le Verrier a avancée. 

En particulier, Benjamin Pierce, professeur à Harvard  communique le 7 décembre 1847 
une nouvelle détermination des éléments elliptiques de Neptune, obtenue par Walker \citep{Pier1848a}.
Ce dernier obtient une excentricité de $0.00857741$, une période de $164.6181$ an et une masse comprise entre $1/19500$ et $1/17000$ masse solaire, 
cette valeur, obtenue grâce aux observations de Triton, étant beaucoup moindre que la valeur de Le Verrier de $1/9300$ masse solaire
\footnote{La masse du système de Neptune adoptée par l'IERS 1992 est de 1/19412.24 masse solaire}. 

Bien que cela nécessite un examen plus approfondi, il est sans doute  normal que Le Verrier, 
ayant supposé une distance au Soleil plus grande, se retrouve aussi avec une masse de Neptune 
plus importante. Les perturbations  principales sont en effet des termes de marées, 
proportionnels à la masse, et inversement proportionnels au cube de la 
distance. Mais cela ne désarme pas la critique.

L'attaque principale viendra de Jacques Babinet, membre de la section de physique de l'Académie des sciences.
Lors de la séance du 21 août 1848  il tient pour acquis 
que plus personne ne croit que la planète calculée par Le Verrier est la même que celle 
observée par Galle 

\bqs
L'identité de la planète Neptune avec la planète théorique, qui rend
compte si admirablement des perturbations d'Uranus, d'après les travaux de
MM. Le Verrier et Adams, mais surtout d'après ceux de l'astronome français,
n'étant plus admise par personne depuis les énormes différences constatées
entre l'astre réel et l'astre théorique quant à la masse, à la durée de la révolution,
à la distance au Soleil, à l'excentricité,\dots \citep{Babi1848a}
\eq

Babinet propose même d'expliquer les différences par l'existence d'une planète additionnelle 
qu'il nomme déjà {\it Hyperion} et qu'il situe au-delà de Neptune \citep{Babi1848a}. 
Le Verrier répond immédiatement à ces attaques \citep{Le-V1848a}, et prolongera 
son argumentation dans un mémoire plus complet dans la séance du 11 septembre 1848 \citep{Le-V1848f}.
A juste titre, Le Verrier explique que sa théorie rend bien compte des positions de Neptune, 
sur une période de 65 ans, sur laquelle on dispose d'observations, 

\bqs
Je détermine, ai-je dit, la position de Neptune au moyen des perturbations
qu'il produit sur Uranus. Quand il y a des perturbations, je puis dire
où est Neptune : mais me demander de le faire longtemps après que l'action
perturbatrice a disparu, c'est tout simplement exiger de moi l'impossible,
une sorte de miracle.

Or, en examinant ma carte, que je mettrai avant peu de jours à la
disposition du public, et sur laquelle j'ai tracé la route d'Uranus, on voit
clairement que cette planète n'a été influencée par l'action de Neptune que
depuis 1812 jusqu'en 1842, c'est-à-dire pendant 3o ans seulement \citep{Le-V1848f}.
\eq

Il poursuit, comme il a été rappelé ci-dessus

\bqs
 La direction était encore plus précise que la distance. Cela devait être,
parce que si la direction eût été fausse, rien n'eût pu compenser l'erreur
qui en fût résultée dans l'attraction que Neptune exerce sur Uranus. Tandis
que si l'on place la planète un peu trop loin dans une direction donnée, on
peut détruire immédiatement l'erreur qui en résulterait sur la quantité de
l'attraction, en faisant la planète un peu plus grosse. C'est précisément ce qui
a eu lieu. J'ai placé Neptune un peu trop loin; mais je l'ai fait un peu trop
gros \citep{Le-V1848f}.
\eq

Le Verrier complètera néanmoins sa démonstration dans la  séance du 2 octobre 1848 \citep{Le-V1848b}.
Il y décrit les incertitudes résultant des différentes hypothèses qu'il a dû prendre, 
tout en montrant que cela n'entachait pas le résultat final d'une erreur 
tellement importante qu'elle aurait interdit la découverte de la planète. 
Il conclut finalement \citep{Le-V1848b}.

\bqs
Qu'on me permette de le dire avec franchise. Lorsque j'annonçai mon
principal résultat en 1846, je ne trouvai presque personne qui voulût y
croire. Déduire la position d'une planète d'un petit dérangement qu'elle
produit sur Uranus! Quelle folie! disait-on. Or ce sont précisément ceux
qui parlaient ainsi qui, aujourd'hui, trouvent tout à fait intolérable que j'aie
réussi à donner la position de Neptune pendant 80 ans sans erreur de plus
de sept degrés et demi aux extrémités de cette période, et qui pensent
qu'on en doit faire un sévère exemple ! 
\eq

Ses arguments sont convaincants, comme en témoigne cette lettre de John Russel Hind, 
astronome britannique,  à Le Verrier 
du 25 septembre 1848 publiée dans les CRAS du 2 octobre 1848

\bqs
J'ai lu avec le plus grand intérêt votre réponse sur Neptune. Mon opinion
sera, je le suppose, la même que celle de toute personne qui a quelque
prétention à une connaissance de l'astronomie; savoir, que vous avez complètement
renversé tous les arguments élevés contre vous ....
\eq

Comme le suppose Hind, cet avis sera largement partagé par la communauté astronomique 
et les honneurs  vont affluer vers Le Verrier qui sera, en particulier, nommé directeur de l'Observatoire de 
Paris en 1854 après le décès d'Arago. 

\section{Retour sur la théorie de Mercure et épilogue.}

Une fois à la direction de l'Observatoire de Paris, Le Verrier entreprend une grande \oe uvre qui consiste dans une refonte 
complète de toutes les solutions  des mouvements  (les "théories") des planètes. 
Son but n'est pas simplement de prédire le plus précisément possible le mouvement des astres, mais bien 
de rechercher s'il manque encore quelque chose dans la modélisation du système solaire pour qu'il y ait un accord parfait entre le 
calcul et les observations. Pour cela, il va commencer par établir une solution la plus précise possible du mouvement de la Terre, 
ou plutôt, dans le langage de l'époque une {\it théorie du mouvement du Soleil}.  Celle-ci est indispensable à la réduction des 
observations, tous les observateurs se trouvant sur Terre. 

Il est conscient que ce sont justement les imprécisions de cette théorie du Soleil qui ont limité ses  résultats 
quand il avait entrepris d'étudier le mouvement de Mercure. Maintenant que ceci est réglé, il peut reprendre son travail 
sur le mouvement de Mercure, comme il le décrit dans une lettre à Hervé Faye publiée dans les CRAS du 12 septembre 1859 \citep{Le-V1859b}.

\bqs
La théorie du Soleil une fois mise hors de cause, il devenait possible
de reprendre avec utilité l'étude des mouvements de Mercure. C'est ce travail
dont je désire vous entretenir aujourd'hui.
\eq

Ce qui rend l'étude de Mercure particulièrement intéressante, c'est l'existence des observations de très haute 
précision que sont les transits de la planète devant le Soleil. La mesure des instants des contacts est extrêmement 
dépendante des positions de la Terre et de Mercure, et ont été enregistrées avec précision depuis 150 ans.

\bqs
On possède, depuis 1697 jusqu'en 1848, vingt et une observations de
cette espèce, auxquelles on doit pouvoir satisfaire de la manière la plus
étroite si les inégalités des mouvements de la Terre et de Mercure ont
été bien calculées, et si les valeurs attribuées aux masses perturbatrices
sont exactes \citep{Le-V1859b}.
\eq    

Cependant, des erreurs subsistent de plusieurs minutes, que Le Verrier ne peut pas croire être 
dues aux observateurs expérimentés que sont Lalande, Cassini, ou Bouguer. Mais il poursuit

\bqs
Mais, ce qui est remarquable, c'est qu'il a suffi d'augmenter de 38 secondes
le mouvement séculaire du périhélie pour représenter toutes les
observations des passages à moins d'une seconde près, et même la plupart
d'entre elles à moins d'une demi-seconde.
\eq 

Ces 38 secondes, il faut les comprendre comme un excès de 38 secondes d'arc par siècle par rapport au mouvement calculé du 
périhélie de Mercure sous l'influence de l'ensemble des autres planètes du système solaire 
qu'il trouve être de 526.7 "/siècle \citep{Le-V1859a}. Avec ces 38"/siècle,  l'ajustement aux observations 
devient si parfait que Le Verrier ne peut les mettre en doute. Il en cherche alors la cause possible. 
Une solution pourrait être d'augmenter la masse de Vénus d'un dixième de sa valeur. Mais augmenter d'une 
telle  valeur la masse de Vénus conduirait à un changement trop important des variations  de l'obliquité de la Terre. 
Il en conclut 

\bqs
Si, au contraire,
on regarde la variation de l'obliquité et les causes qui la produisent
comme bien établies, on sera conduit à penser que l'excès du périhélie du 
mouvement de Mercure est dû à quelque action encore inconnue.
\eq

Le Verrier continue en disant qu'un tel effet pourrait être dû à l'action d'une planète 
intérieure  à l'orbite de Mercure. Comme il s'étonne qu'alors personne ne l'ait observée 
lors d'un passage sur le disque du Soleil, il envisage la possibilité d'une multitude de corps plus petits 
qui deviendraient alors très difficiles à observer. 
Hervé Faye prend très au sérieux les observations de Le Verrier, et dans une réponse publiée dans le même 
numero des CRAS, il propose à l'Académie de mettre en place une stratégie d'observation pour rechercher 
ces planètes intra mercurielles \citep{Faye1859a}.

Le Verrier va longtemps chercher cette planète qu'il dénomme Vulcain.  Quand, le 22 décembre 1859,  
il reçoit une lettre d'un astronome amateur, 
le Dr. Edmond Lescarbault, habitant de Orgère-en-Beauce faisant état du passage possible d'une planète devant  le Soleil, 
Le Verrier est particulièrement  intéressé et en fait état lors de la séance de l'Académie du  2 janvier 1860 \citep{Lesc1860a}.
Avec force précision, le Dr. Lescarbault décrit les observations qu'il a consignées, mentionnant le passage, le 26 mars 1859, 
d'un corps passant devant le Soleil. 
Le Verrier fait part de ses remarques à la suite du texte de la lettre \citep{Le-V1860h}. Sa curiosité 
l'avait poussé à se rendre directement chez le Dr. Lescarbault.

\bqs
On pouvait être surpris toutefois que
M. Lescarbault, se trouvant en possession d'un fait aussi considérable, fût
demeuré neuf mois sans en donner connaissance. Cette considération m'a
déterminé à me rendre sur-le-champ à Orgères, où M. Vallée fils, ingénieur
des Ponts et Chaussées, a bien voulu m'accompagner, et où nous sommes
arrivés le samedi 31 décembre sans avoir été annoncés.
Nous avons trouvé en M. Lescarbault un homme adonné depuis longtemps
à l'étude de la science, entouré d'instruments, d'appareils de toute
nature, construisant lui-même et ayant fait édifier une petite coupole tournante.
M. Lescarbault a bien voulu nous permettre d'examiner dans le plus
scrupuleux détail les instruments dont il s'est servi, et il nous a donné les
explications les plus minutieuses sur ses travaux et en particulier sur toutes
les circonstances du passage d'une planète sur le Soleil \citep{Le-V1860h}.
\eq

On le sait maintenant, il n'y avait point de planète Vulcain, que Le Verrier continuera 
de chercher en vain. En revanche les 38 secondes par siècle d'excès du mouvement du périhélie de Mercure 
étaient bien réelles, 
mais pour les expliquer, il fallait effectuer ce  Le Verrier ne voulait se 
résoudre à considérer qu'en dernière extrémité, après avoir exclu   toutes les 
autres hypothèses 

\bqs
L'altération des lois de la gravitation
serait une dernière ressource à laquelle il ne pourrait être permis
d'avoir recours qu'après avoir épuisé l'examen des autres causes, qu'après les
avoir reconnues impuissantes à produire les effets observés \citep{Le-V1846d}.
\eq

En faisant  état de cette avance de 38"/siècle qu'il avait clairement identifiée, au prix d'un 
travail phénoménal, Le Verrier rendait un service considérable à la science. Cet écart sera en effet 
la  première  vérification de la relativité générale d'Einstein, publiée en 1915, qui donne une avance 
du périhélie de Mercure de 43"/siècle, très proche de la valeur calculée de Le Verrier. 
Il fallait donc bien se résoudre cette fois-ci à changer la loi de gravitation. 
A l'heure actuelle, des calculs similaires à ceux de Le Verrier, mais exécutés numériquement 
sur ordinateurs, en tenant compte des données des sondes spatiales orbitant autour des planètes, 
permettent de vérifier que la loi de gravitation fournie par la relativité générale d'Einstein 
représente bien les avances des périhélies des planètes à une précision relative de quelques  
$10^{-5}$ \citep[e.g.][]{Fien2015a}.

\section*{Remerciements}
L'auteur remercie J. Lequeux pour ses éclaircissements sur le fonctionnement de l'Académie des sciences 
à l'époque d'Arago, ainsi que A. Albouy, G. Bertrand, N. Capitaine, A. Chenciner, D. Briot, et M. Postel pour leurs 
relectures de ce texte. Ce travail doit beaucoup à la numérisation des Comptes Rendus de l'Académie des sciences, 
disponibles librement sur (gallica.bnf.fr), ainsi qu'à la numérisation des procès verbaux 
des séances du Bureau des longitudes sur (bdl.ahp-numerique.fr).



\bibliographystyle{apalike}  
\bibliography{leverrier} 




\end{document}